\documentclass[aps,twocolumn,pra]{revtex4}
\usepackage{graphicx}
\DeclareGraphicsExtensions{.pdf}

\newcommand{\beq}{\begin{equation}}
\newcommand{\eeq}{\end{equation}}
\newcommand{\bea}{\begin{eqnarray}}
\newcommand{\eea}{\end{eqnarray}}

\newcommand{\bra}[1]{\left< #1 \right|}
\newcommand{\ket}[1]{\left| #1 \right>}

\newcommand{\eu}{\mathrm{e}}
\newcommand{\tr}{\mathrm{tr}}

\newcommand{\st}{^\star}
\newcommand{\Hc}{\mathrm{H.c.}}

\newcommand{\half}{\frac{1}{2}}

\newcommand{\Xo}{{\hat X}}
\newcommand{\Ao}{{\hat A}}
\newcommand{\xo}{{\hat x}}
\newcommand{\Ho}{{\hat H}}
\newcommand{\qo}{{\hat q}}
\newcommand{\po}{{\hat p}}
\newcommand{\co}{{\hat c}}
\newcommand{\cod}{{\hat c^\dagger}}

\newcommand{\Uo}{{\hat U}}
\newcommand{\phio}{{\hat\varphi}}
\newcommand{\Fo}{{\hat F}}
\newcommand{\Tim}{\mathcal{T}}
\newcommand{\Nor}{\mathcal{N}}
\newcommand{\Wey}{\mathcal{W}}
\newcommand{\Ord}{\mathcal{O}}
\newcommand{\Per}{\mathcal{P}}

\newcommand{\al}{\alpha}
\newcommand{\be}{\beta}
\newcommand{\la}{\lambda}
\newcommand{\eps}{\epsilon}
\newcommand{\om}{\omega}
\newcommand{\si}{\sigma}
\newcommand{\pr}{^\prime}
\newcommand{\ka}{\kappa}

\begin{document}

\title{Wick theorem for all orderings of canonical operators}
\author{Lajos Di\'osi}
\affiliation{Wigner Research Centre for Physics, 
                      H-1525 Budapest 114, P. O. Box 49, Hungary}
\begin{abstract}
Wick's theorem, known for  yielding normal ordered from time-ordered bosonic fields, 
may be generalized for a simple relationship between any two orderings that we
define over canonical variables, in a broader sense than before. 
In this broad class of orderings, the General Wick Theorem (GWT) follows 
from the Baker--Campbell--Hausdorff identity. 
We point out that, generally, the characteristic function does not induce an unambigous
scheme to order the multiple products of the canonical operators although the
value of the ordered product is unique.  We construct a manifold of different schemes 
for each value of $s$ of s-orderings of Cahill and Glauber.
\end{abstract}


\maketitle

\section{Introduction}
The problem of systematic ordering of canonical operators $\qo,\po$ appeared first in  
canonical quantization \cite{Wey27} and also the other way around: in 
classical phase-space representations of canonically quantized systems \cite{Wig32}.
In a different realm, ordering and reordering of quantized fields became central
to S-matrix theory \cite{HouKin49,Dys49,Wic50}.

A particular phase-space quasi-distribution $\rho(q,p)$ 
for a given density matrix $\hat\rho$, 
\begin{equation}
\rho(q,p)=\tr[\Ord\delta(q-\qo)\delta(p-\po)\hat\rho],
\end{equation}
requires a particular ordering $\Ord$ of $\qo$ and $\po$.
The Weyl--Wigner \cite{Wey27,Wig32}, normal \cite{Wic50} and 
anti-normal orderings are special cases of s-orderings proposed by 
Cahill and Glauber \cite{CahGla69}, 
reviewed together with QP- and PQ-orderings e.g. in \cite{Lee95}. 

An ordering $\Ord$ is traditionally defined by its action
on the operator-valued characteristic function. 
In particular, the Weyl--Wigner ordering $\Wey$ is defined by this relationship:
\begin{equation}\label{Wey}
\Wey\eu^{a\qo+b\po}=\eu^{a\qo+b\po},
\end{equation}
with the arbitrary c-numbers $a,b$. 
The normal ordering $\Nor$ is defined between the annihilation  $\co=(\xo+i\po)/\sqrt{2}$ 
and creation $\cod$ operators  (we use units where $\hbar=1)$:  
\beq
\Nor\eu^{\la\st\co+\la\cod}=\eu^{\la\cod}\eu^{\la\st\co},
\eeq
with an arbitrary  complex number $\la$.
The s-orderings are defined by
\beq\label{Ord_s}
\Ord_s\eu^{\la\st\co+\la\cod}=\eu^{-s\vert\la\vert^2/2}\eu^{\la\cod+\la\st\co},
~~~~~s\in[-1,1].
\eeq
They interpolate between the normal ordering $(s=1)$ and
anti-normal ordering $(s=-1)$, with the Weyl--Wigner ordering in the middle $(s=0)$.

Time-ordering $\Tim$ (also called chronological or path-ordering) was proposed first by
Dyson \cite{Dys49}  in relativistic quantum field theory.  
The S-matrix is a time-ordered functional of free quantum field operators. 
If we transform it into the normal ordered form
first, the evaluation of matrix elements becomes straightforward. 
Time- and normal orderings are related by Wick's theorem \cite{Wic50}. 
For a free relativistic bosonic field $\phio(x)$ it can be written into the compact form, cf. \cite{Choetal85}:
\beq
\Tim\eu^{\int J(x)\phio(x)dx}=\eu^{C}\Nor\eu^{\int J(x)\phio(x)dx},
\eeq
where $J(x)$ is an arbitrary c-number field. 
The exponent in the pre-factor is a c-number:
\beq
C=\int\int J(x)C(x,y)J(y)dxdy,
\eeq
where $C(x,y)$ is called the kernel of chronological \emph{contraction} (or pairing):
\beq
C(x,y)=\Tim\phio(x)\phio(y)-\Nor\phio(x)\phio(y)\equiv(\Tim-\Nor)\phio(x)\phio(y).
\eeq
The above form of Wick's theorem suggests
that a generalized theorem of the same structure holds for any pair $\Ord$
and $\Ord\pr$ of orderings.  

In Sec. \ref{Monord} we define the notion of monomial and non-monomial orderings and we note that, in the latter class,
the ordered characteristic function allows for ambiguous schemes of ordering. Sec. \ref{GWT} suggests 
the general Wick theorem for monomial orderings, with a tentative proof in Sec. \ref{Proof}.
Sec. \ref{Exam} illustrates the theorem on simple applications in quantum mechanics and optics.
Sec. \ref{s_Ord} proves that an s-ordering of Cahill and Glauber is equivalent with a family of `redundant' monomial
path orderings, and we construct the corresponding family of ordering schemes.  

\section{Monomial and non-monomial orderings}\label{Monord}
Let us start with a collection $\{\Ao_\al, \al\in\Omega\}$ of operators to be ordered, where $\Omega$
is a (partially) ordered set of labels. (If  $[\Ao_\al,\Ao_\be]=0$ for a certain pair $\al,\be\in\Omega$
then  the order of $\al$ and $\be$ may be left unspecified.) Consider an ordered subset 
$\{\al_1,\al_2,\dots\al_n\}=\Omega\pr\subseteq\Omega$ of labels 
and define the corresponding ordered operator product: 
\bea\label{perm}
\Ord\prod_{\al\in\Omega\pr}\Ao_\al&=&\Ao_{\al_n}\dots\Ao_{\al_3}\Ao_{\al_2}\Ao_{\al_1}\nonumber\\
&\equiv&\left[\prod_{\al\in\Omega\pr}\Ao_\al\right]_P
\eea
where $\Ord$ stands for the given operator ordering to adopt the ordering $\al_n\succ\dots \succ\al_3\succ\al_2\succ\al_1$
of the labels; $[\dots]_P$ is alternative notation referring to the corresponding permutation $P$. 
The collection of the operators may be redundant:  a certain $\Ao_\al$ may coincide with 
a certain $\Ao_\be$ even for $\al\neq\be$.   Let us call (\ref{perm}) monomial ordering (or permutation).
Ordering, in general, can be the  weighted mixture of different monomial orderings:
\beq\label{mix}
\Ord\prod_{\al\in\Omega\pr}\Ao_\al=\sum_P w_P \left[\prod_{\al\in\Omega\pr}\Ao_\al\right]_P
\eeq
where $w_P$ are non-negative weights that sum up to $1$. 
 
Alternatively, orderings used to be defined by their exponential characteristic functions
\beq
\Fo_\Ord(\la)=\Ord\!\!\prod_{\al\in\Omega}\!\!\eu^{\la_\al\Ao_\al}
\eeq
at arbitrary arguments $\la_\al$. 
For monomial orderings (\ref{perm}) they factorize: 
\beq\label{ord}
\Fo_\Ord(\la)
=\dots\exp(\la_{\al_3}\Ao_{\al_3})\exp(\la_{\al_2}\Ao_{\al_2})\exp(\la_{\al_1}\Ao_{\al_1}),
\eeq
where  $\dots\succ\al_3\succ\al_2\succ\al_1$ holds. As we see, the \emph{value} of the characteristic function
determines  the \emph{scheme} (\ref{perm}) of monomial ordering (and vice versa).
This is not so for non-monomial orderings. The characteristic function does not determine
the scheme (\ref{mix}) of ordering uniquely. 

Typical non-monomial ordering is the Weyl--Wigner ordering $\Wey$ defined by its characteristic function (\ref{Wey}).
It turns a monomial into weighted sum of ordered monomials. For instance:
\beq
\Wey\qo^2\po=\frac{\qo^2\po+\po\qo^2+\qo\po\qo}{3}
                             = \frac{\qo^2\po+\po\qo^2+2\qo\po\qo}{4}
\eeq
where we show two (of the many) schemes of weighted permutations (\ref{mix}) which correspond to the same non-monomial
ordering $\Wey$. Still, we anticipate a trick (from Sec. \ref{s_Ord}) to show that the Weyl--Wigner ordering can be
squeezed into the class of monomial orderings. The trick is that we introduce the redundant collection  
$\{\qo_\tau,\po_\tau;\tau\in[0,1]\}$ of the canonical operators, where $\qo_\tau=\qo$ and
$\po_\tau=\po$ for all $\tau\in[0,1]$., i.e.: we assign the time-label $\tau$ formally. Then time-ordering, which
is typical monomial ordering, leads to Weyl--Wigner ordering:
\bea\label{Wey-Tim}
\Tim\!\exp\!\left(\int_0^1\!\!\!(a\qo_\tau+b\po_\tau)d\tau\!\right)\!\!\!
&=&\!\!\!\lim_{\eps\rightarrow+0}\!\left(\eu^{\eps(a\qo+b\po)}\right)^{[1/\eps]}\!\!\!=\eu^{a\qo+b\po}=\nonumber\\
&=&\Wey\eu^{a\qo+b\po}.
\eea
This is simplest redefinition of $\Wey$ as a particular monomial time-ordering, there are infinite many
other choices which we shall detail and extend for all s-orderings in Sec. \ref{s_Ord}. 

We state our general Wick's theorem (Sec. \ref{GWT}) for monomial orderings of canonical operators,
and construct a tentative proof in Sec. \ref{Proof}. 
The theorem can not be extended for non-monomial orderings in general.
There is, however, a backdoor for some of them, like the above example of Weyl--Wigner ordering, and
all s-orderings, cf. in Sec. \ref{s_Ord}.

\section{General Wick Theorem}\label{GWT}
Consider a canonical system of $n$ pairs of canonical variables 
$\xo_k=\qo_k$ and $\xo_{n+k}=\po_k$ respectively,
where $[\qo_k,\po_l]=i$  for $k,l=1,2,\dots,n$.
We introduce the linear combinations of the canonical variables:
\beq\label{all_lin}
\Xo=\sum_{k=1}^{2n}a_k\xo_k, 
\eeq
with arbitrary coefficients $\{a_k\}$.
Consider a collection of linear combinations
\beq\label{opers}
\Ao_\al=\sum_{k=1}^{2n}A_{\al k}\xo_k,~~~~~\al\in\Omega,
\eeq
which form a complete (or overcomplete) basis in the linear space $\{\Xo\}$.   
Let them be the operators to be ordered as given in Sec. \ref{Monord}. 
If $\dots\succ\al_3\succ\al_2\succ\al_1$ holds then
\bea\label{ord}
\Ord\eu^\Xo &=&\Ord\prod_{\al\in\Omega}\eu^{\la_\al\Ao_\al}\\
&=&\dots\exp(\la_{\al_3}\Ao_{\al_3})\exp(\la_{\al_2}\Ao_{\al_2})\exp(\la_{\al_1}\Ao_{\al_1}),\nonumber
\eea
where 
\beq\label {decomp}
\Xo=\sum_{\al\in\Omega}\la_\al\Ao_\al.
\eeq
It is important to anticipate that whenever we write $\Ord\eu^{\Xo}$ or  $\Ord\Xo^2$,
we understand the above  linear combination in terms of 
the operators to be ordered.  If $\{\Ao_\al\}$ forms an overcomplete basis, 
it matters for the ordering that the coefficients $\{\la_\al\}$ are not unique. In the case of simple
redundancy $\Ao_\al=\Ao_\be$ at $\al\neq\be$ we can, if we wish to, remove this ambiguity by collapsing
the coefficients: $\la_\al=\la_\be$.

To prepare our general Wick theorem (GWT), 
we consider another collection of operators to be ordered another way $\Ord\pr$:
\beq\label{opers1}
\Ao_a=\sum_{k=1}^{2n}A_{a k}\xo_k,~~~~~a\in O.
\eeq
If  $\dots\succ a_3\succ a_2\succ a_1$ holds then
\bea\label{ord1}
\Ord\pr\eu^\Xo&=&\Ord\pr\prod_{a\in O}\eu^{\la_a\Ao_a}\\
&=&\dots\exp(\la_{a_3}\Ao_{a_3})\exp(\la_{a_2}\Ao_{a_2})\exp(\la_{a_1}\Ao_{a_1}),\nonumber
\eea
where 
\beq\label {decomp1}
\Xo=\sum_{a\in O}\la_a\Ao_a.
\eeq

We propose the following relationship between the two orderings:
\beq\label{GWT1}
\Ord\pr\eu^\Xo = \eu^C\Ord\eu^\Xo,
\eeq
where the pre-factor is a c-number because the exponent  is c-number:
\beq\label{GWT2}
C=\half(\Ord\pr-\Ord)\Xo^2.
\eeq
We call it the \emph{general contraction} between $\Ord\pr$ and $\Ord$.
This is the GWT, our central result. 

If we substitute $\Ord\pr=\Wey$ and the  identity $\Wey\Xo^2=\Xo^2$, 
we get a simple equivalent form of GWT:
\bea\label{GWT3}
\Ord\eu^\Xo&=&\eu^C\eu^\Xo\label{GWT3a}\\
C&=&\half(\Ord\Xo^2-\Xo^2),\label{GWT3b}
\eea
where $C$ is a c-number, being the contraction between $\Ord$ and $\Wey$.
To express its detailed structure, we insert the decomposition (\ref{decomp}) of $\Xo$:
\bea
C&=&\half\sum_{\al,\be}\la_\al\la_\be
\left(\Ord\Ao_\al\Ao_\be-\half\{\Ao_\al,\Ao_\be\}\right)\nonumber\\
&=&\half\sum_{\al\succ\be}\la_\al\la_\be[\Ao_\al,\Ao_\be]. 
\eea
To further simplify the result, we use the expansion (\ref{opers}) of $\Ao_\al$,
with the new notations $B_{\al k}=A_{\al,n+k}$ for the coefficients of the momenta, 
yielding:
\beq
C=\frac{i}{2}\sum_{\al\succ\be}
\la_\al\la_\be\sum_{k=1}^n(A_{\al k}B_{\be k}-A_{\be k}B_{\al k}).
\eeq

We note that in the special case when
$\Ord\pr$ and $\Ord$ are to order the same collection of operators (in
two different ways, of course) then GWT (\ref{GWT1}-\ref{GWT2}) can be stated for the characteristic
functions as well:
\bea
\Fo_{\Ord\pr}(\la)&=&\exp\left(\half\sum_{\al,\be}C_{\al\be}\la_\al\la_\be\right)\Fo_\Ord(\la)\\
C_{\al\be}&=&\frac{\partial^2}{\partial\la_\al\partial\la_\be}\left[\Fo_{\Ord\pr}(\la)-\Fo_{\Ord}(\la)\right]_{\la=0}.\nonumber
\eea

\section{Tentative proof}\label{Proof}
Let us consider a given $\Xo$, as in (\ref{all_lin}), and its decompositions 
(\ref{decomp}) and (\ref{decomp1}):
\beq
\Xo=\sum_{k=1}^{2n}a_k\xo_k =\sum_{\al\in\Omega}\la_\al\Ao_\al 
=\sum_{a\in O}\la_a\Ao_a.
\eeq
The proof of GWT (\ref{GWT1}-\ref{GWT2}) simplifies if we restrict
ourselves for the case when  all coefficients $a_k,A_{\al k},A_{a k}$ 
are non-negative. Then we construct the asymptotic form of the 
maximally refined decomposition of $\Xo$:
\beq\label{decomp_maxref}
\Xo=\lim_{\eps\rightarrow+0}\sum_{k=1}^{2n}
(~\underbrace{\eps\xo_k+\!\!\eps\xo_k+\dots+\!\!\eps\xo_k}_{[a_k\!/\!\eps]}~)
\eeq
If we approach the limit $\eps\rightarrow0$, the two orderings in question act on the \emph{same} (yet unordered) product of
exponentials:
\beq\label{ord_maxref}
\begin{array}{lcl}
\Ord\eu^\Xo&=&\lim_{\eps\rightarrow+0}\Ord\\
\Ord\pr\eu^\Xo&=&\lim_{\eps\rightarrow+0}\Ord'
\end{array}
\left\{\prod_{k=1}^{2n}(~\underbrace{\eu^{\eps\xo_k}\times\eu^{\eps\xo_k}\times\dots\times\eu^{\eps\xo_k}}_{[a_k\!/\!\eps]}~)\right.,
\eeq
where the number of factors $\eu^{\eps\xo_k}$ is $[a_k/\eps]$ for all $k$.
It will be the order of the factors, and nothing else, that distinguishes $\Ord$ and $\Ord\pr$.
From one to the other, we can go by a finite
sequence $P_1,P_2,\dots,P_K$ of $K$ switches between neighboring factors. 
($K\rightarrow\infty$ for $\eps\rightarrow+0$.)
Namely,
\beq\label{perms}
\Ord\pr\eu^\Xo=\Per_K\dots\Per_2\Per_1\Ord\eu^\Xo,
\eeq
where we understand the same (redundant) factorization
(\ref{ord_maxref}) for $\eu^\Xo$.
Let us consider the r.h.s. of the above equation and 
suppose $\Per_K$ inverts the order of neighbors $\eu^{\eps\xo_k}\eu^{\epsilon\xo_l}$ in
$\Per_{K-1}\dots\Per_1\Ord\eu^\Xo$:
\beq
\Per_K\eu^{\eps\xo_k}\eu^{\epsilon\xo_l}=\eu^{\eps\xo_l}\eu^{\epsilon\xo_k}
\equiv\eu^{\eps^2[\xo_l,\xo_k]}\eu^{\eps\xo_k}\eu^{\epsilon\xo_l},
\eeq
where we used the simple Baker--Campbell--Hausdorff \cite{BCH} identity.
We can write:
\beq\label{P1}
\Per_K\dots\Per_1\Ord\eu^\Xo=\eu^{\eps^2[\xo_l,\xo_k]}\Per_{K-1}\dots\Per_1\Ord\eu^\Xo
\eeq
Observe that
\beq\label{P1a}
\Per_K\dots\Per_1\Ord\Xo^2-\Per_{K-1}\dots\Per_1\Ord\Xo^2=2\eps^2[\xo_l,\xo_k].
\eeq
Using this,  the r.h.s. of (\ref{P1}) reads:
\beq\label{P1b}
\eu^{\half(\Per_K\dots\Per_1-\Per_{K-1}\dots\Per_1)\Ord\Xo^2}
\Per_{K-1}\dots\Per_1\Ord\eu^\Xo
\eeq
We  repeat the same identity transformation on the factor $\Per_{K-1}\dots\Per_1\Ord\eu^\Xo$,
yielding
\beq\label{P2}
\eu^{\half(\Per_K\dots\Per_1-\Per_{K-2}\dots\Per_1)\Ord\Xo^2}
\Per_{K-2}\dots\Per_1\Ord\eu^\Xo.
\eeq
After $K$ steps, we get
\beq\
\eu^{\half(\Per_K\dots\Per_1\Ord-\Ord)\Xo^2}\Ord\eu^\Xo=\eu^{\half(\Ord\pr-\Ord)\Xo^2}\Ord\eu^\Xo
\eeq
where we used $\Ord\pr=\Per_K\dots\Per_1\Ord$ from (\ref{perms}). This is our ultimate expression
for the r.h.s. of (\ref{GWT1}). Our GWT is confirmed.

Recall that we restricted the proof: the operators 
to be ordered had to be linear combinations of the canonical variables with 
non-negative coefficients. The maximally refined decomposition (\ref{decomp_maxref}) contained
terms with the unique positive coefficient $\epsilon$. If we want to extend
the proof for orderings of operators without the above restriction, we have to construct
the maximally refined decomposition for all $\Xo$.
There will be four types of coefficients:
$\pm\epsilon,\pm i\epsilon$ instead of the positive ones $\epsilon$.
Once we have constructed such a maximally refined decomposition to host both
$\Ord$ and $\Ord\pr$, the proof follows the same steps as before --- and needs a definitely more
intricate book-keeping of the four types of exponentials. Intuition says, nonetheless,
that the validity of our theorem extends from positive parameters for all 
complex ones just by the theorem's analytic form.

\section{Examples: $\Tim$ versus QP versus $\Nor$ orderings}\label{Exam}
We are going to illustrate the flexibility of GWT in solving simple tasks
in elementary quantum mechanics and quantum optics. 

Consider the 
Schr\"odinger equation $\dot\psi_t=-\Ho_t\psi_t$ of a mass $m$,
where the Hamiltonian
$\Ho_t=(\po^2/2m)-F_t\qo$
contains a given time-dependent force.
Let us find the solution $\psi_t^{I}=\Uo_t\psi_0$ 
in the interaction picture where $\qo_t=\qo+\po t/m$ and $\po_t=\po$.
(As is known, the solution $\psi_t$ in Schr\"odinger picture is obtained if we solve the
force-free Schr\"odinger equation  with $\psi_t^{I}$ as the initial state.)
The evolution operator in interaction picture is a $\Tim$-ordered exponential:
\beq\label{U_t}
\Uo_t=\Tim\eu^{\Xo_t},
\eeq
where $\Xo_t$ is linear combination of the canonical variables:  
\beq\label{Sch_tpart}
\Xo_t=i\int_0^t F_\tau \qo_\tau d\tau.
\eeq
The r.h.s. is the \emph{chronological} decomposition of $\Xo_t$: 
the summation label $\alpha\in\Omega$ in (\ref{decomp}) became 
the integral variable $\tau\in(0,t)$, the set  $\{\Ao_\al\}$ became  $\{\qo_\tau\}$,
the coefficients $\la_\al$  became $iF_\tau$.
To evaluate $\Uo_t$, we shall consider its QP-ordering, hence we need a
\emph{canonical} decomposition of $\Xo_t$:
\beq\label{Sch_canpart}
\Xo_t=i\int_0^t F_\tau d\tau \qo +\frac{i}{m}\int_0^t F_\tau \tau d\tau \po
\equiv i\Delta p_t\qo - i\Delta q_t\po,
\eeq
where we introduced the momentum and coordinate shifts $\Delta p_t,\Delta q_t$ 
for notational convenience.  The label $a$ in (\ref{decomp1})  takes two values only: 
$\Ao_{a=1}=\qo$ and  $\Ao_{a=2}=\po$,
also $\la_{a=1}=i\Delta p_t$ and  $\la_{a=2}=-i\Delta q_t$.
The QP-ordering $\Ord_{QP}$ pushes all $\qo$ to the left of all $\po$. 
Now we apply our GWT (\ref{GWT1}) to (\ref{U_t}),   
\beq
\Uo_t=\Tim\eu^{\Xo_t}=\eu^{C_t}\eu^{i\Delta p_t\qo}\eu^{-i\Delta q_t\po},
\eeq
with the contraction (\ref{GWT2}):
\beq\label{Sch_contr}
C_t\equiv\half(\Tim-\Ord_{QP})\Xo_t^2=
\frac{i}{2}\int_0^t\int_0^t F_\tau F_\si \vert\tau-\si\vert d\tau d\si.
\eeq
$\Uo_t\psi_0$ provides the following explicit solution for the wave function 
in the interaction picture:
\beq\label{psi_t}
\psi_t^I(q)=\eu^{C_t}\eu^{i\Delta p_t q}\psi_0(q-\Delta q_t).
\eeq

A similar standard task is the electromagnetic cavity oscillator of frequency $\om$
under external driving. The Hamiltonian reads:
$\Ho_t=\om\cod\co+(E_t^\ast\co-\Hc)$.
In interaction picture, where $\co_t=\eu^{-i\om t}\co$ and $\cod_t=\eu^{i\om t}\cod$,
the evolution operator is (\ref{U_t}) again, with
\beq
\Xo_t=\int_0^t (E_\tau^\ast\co_\tau-\Hc)d\tau
       \equiv\Delta c_t\st \co  - \Hc   
\eeq
We can rewrite the evolution operator $\Uo_t$ into the normal
ordered form, applying the GWT (\ref{GWT1}-\ref{GWT2}):
\beq
\Uo_t=\eu^{\half(\Tim-\Nor)\Xo_t^2}\Nor\eu^{\Xo_t}
=\eu^{C_t}\eu^{-\Delta c_t\cod}\eu^{\Delta c_t\st\co},
\eeq
\beq
C_t=-\int_0^t\int_0^t \theta(\tau-\si)E^\ast_\tau E_\si \eu^{-i\om(\tau-\si)} d\tau d\si.
\eeq

In quantum optics, there is a spectacularly simple definition of the squeezing operator
with squeezing parameter $\mu\rangle0$: 
\beq\label{S1}
{\hat S}=\frac{1}{\sqrt{\mu}}\int \ket{q/\mu}\bra{q}dq=\Ord_{PQ}\eu^{i(1-1/\mu)\po\qo},
\eeq
where the first expression was proposed in Refs.  \cite{Fanetal} while the second compact
form is our finding based on it.  To achieve the normal ordered form, which was the task in
Refs.  \cite{Fanetal} as well, we first unravel the 
bilinear form $\po\qo$ in the exponent. Introduce $\ka=1-1/\mu$ and consider
\beq\label{S2}
{\hat S}=\Ord_{PQ}\eu^{i\ka\po\qo}=\int\Ord_{PQ}\eu^{iz\po\pm z\st\qo}\exp\left(-\frac{\vert z\vert^2}{\vert\ka\vert}\right)\frac{d^2z}{\pi\vert\ka\vert},
\eeq
where the sign $\pm$ is the sign of $\ka$. Now we can apply our GWT (\ref{GWT1}-\ref{GWT2}):
\beq\label{S3}
\Ord_{PQ}\eu^{iz\po\pm z\st\qo}=\eu^{\frac{1}{4}(z\st)^2-\frac{1}{4}z^2\pm\half\vert z\vert^2}\Nor\eu^{iz\po\pm z\st\qo}.
\eeq
Inserting this in (\ref{S2}) and performing the complex integral, we obtain:  
\beq\label{S4}
\begin{array}{lcl}
{\hat S}\!\!\!&=&\!\!\!\sqrt{\!\frac{2\mu}{1\!+\!\mu^2}}\Nor\!\exp\!\left\{\frac{-i(1\!-\!\mu^2)\po\qo\!-\!\half(1\!-\!\mu)^2(\po^2\!+\!\qo^2)}{1+\mu^2}\right\}=\\
                       &=&\!\!\!\sqrt{\!\frac{2\mu}{1\!+\!\mu^2}}\Nor\!\exp\left\{\frac{-\half(1\!-\!\mu^2)(\co^2\!-\!\co^{\dagger2})\!+\!(1\!-\!\mu)^2\cod\co}{1+\mu^2}\right\}.
\end{array}
\eeq

\section{s-Ordering}\label{s_Ord} 
The s-orderings are defined by their characteristic function  (\ref{Ord_s}).
Apart from the two marginal cases $s=\pm1$ they are non-monomial
and  the induced scheme (\ref{mix}) of ordering is not unique. First we 
redefine a given s-ordering as a path (i.e.: monomial) ordering
of a redundant collection of the canonical operators,
then we can construct the corresponding scheme of their s-ordering. 

\subsection{s-Ordering as Path Ordering}
We are going to show that an s-ordering is equivalent with 
path-orderings $\Tim$ of a (redundant) collection of operators
$\{\co_\tau,\cod_\tau; \tau\in[0,1]\}$ where we take
$\co_\tau\equiv\co$ and $\cod_\tau\equiv\cod$ \emph{after} the
path-ordering only.
 To define a concrete ordering, we must first introduce
a decomposition of the exponent $\Xo=\la\st\co+\la\cod$. 
Let us choose the following structure:
\beq\label{part_s}
\Xo=\int_0^1\left(\la\st \co_\tau d\tau +\la\cod_\tau d\chi_\tau\right),
\eeq
where $\chi_0=0$ and $\chi_1=1$, also $\chi_\tau$ must be
monotonous. As we said, we cancel the label $\tau$ of $\co,\cod$ after the path-ordering. 
Along the path, the rate of $\cod$'s versus
the constant rate of $\co$'s is ruled by $\chi_\tau$.
We prove below that this path-ordering is equivalent with s-ordering (\ref{Ord_s}):
\beq\label{s_T_equiv}
\Ord_s\eu^{\la\st \co +\la\cod}=
\left.\Tim\exp
\left\{\int_0^1\left(\la\st \co_\tau d\tau +\la\cod_\tau d\chi_\tau\right)\right\}\right\vert_{\co_\tau=\co},
\eeq
provided $\chi_\tau$ satisfies
\beq\label{s_chi}
s=1-2\int_0^1\!\!\!\! \chi_\tau d\tau.
\eeq

The proof goes like this.
Considering the decomposition (\ref{part_s}) of $\Xo$, we write
the r.h.s. of (\ref{s_T_equiv}) as $\Tim\eu^\Xo$ and apply 
(\ref{GWT3a}-\ref{GWT3b}) to it:  
\beq\label{T_Xchi}
\Tim\eu^\Xo=\eu^C\eu^{\la\st\co+\la\cod},
\eeq
where
\beq\label{contr_s}
C=\half\left(\Tim\Xo^2-\Xo^2\right)=
\vert\la\vert^2\left(\int_0^1\!\!\!\!\chi_\tau d\tau-\half\right).
\eeq
The steps that led to this expression of  $C$ were the following. 
\beq
\Tim\Xo^2=\int_0^1\int_0^1 
\Tim\left(\la\st \co_\tau d\tau +\la\cod_\tau d\chi_\tau\right)
         \left(\la\st \co_\si    d\si     +\la\cod_\si    d\chi_\si    \right).
\eeq
Let us write $\Xo^2$ into the similar integral form:
\beq
\Xo^2=\int_0^1\int_0^1 
\left(\la\st \co d\tau +\la\cod d\chi_\tau\right)
\left(\la\st \co d\si     +\la\cod d\chi_\si   \right).
\eeq
Then we get
\bea
&&2C=\Tim\Xo^2-\Xo^2=\nonumber\\
     &=&\!\!\!\vert\la\vert^2\!\!\!\int_0^1\!\!\int_0^1\!\!\!
              \left\{\left(\Tim \co_\tau\cod_\si\! -\! \co\cod\right)\!d\tau d\chi_\si
                         +\left(\Tim \cod_\tau\co_\si\!-\! \cod\co\right)\!d\chi_\tau d\si\right\}\nonumber\\
&=&\!\!\!\vert\la\vert^2\!\!\!\int_0^1\!\!\int_0^1\!\!\!
              \theta(\si-\tau)(d\chi_\tau d\si-d\tau d\chi_\si)\nonumber\\
&=&\!\!\!\vert\la\vert^2\left(2\int_0^1\!\!\!\!\chi_\si d\si-1\right).
\eea
Therefore (\ref{T_Xchi}), i.e., the r.h.s. of (\ref{s_T_equiv}), reads:
\beq
\Tim\eu^\Xo
=\exp\left(\vert\la\vert^2\int_0^1\!\!\!\!\chi_\tau d\tau-\frac{\vert\la\vert^2}{2}\right)
\eu^{\la\st\co+\la\cod}.
\eeq
This, at the condition (\ref{s_chi}), becomes $\Ord_s\eu^{\la\st\co+\la\cod}$ as defined
by (\ref{Ord_s}).

\subsection{Non-monomial schemes induced by s-ordering}\label{Ord_s_prod} 
Derivatives of the characteristic function (\ref{Ord_s}) yield the s-ordered 
products of the canonical operators $\co,\cod$:
\bea\label{ord_prod_derivs}
\Ord_s\co^n(\cod)^m
&=&\left(\frac{\partial}{\partial \la\st}\right)^n
  \left(\frac{\partial}{\partial \la}\right)^m 
  \left.\Ord_s\eu^{\la\st\co+\la\cod}\right\vert_{\la=0}\\
&=&\left(\frac{\partial}{\partial \la\st}\right)^n
  \left(\frac{\partial}{\partial \la}\right)^m 
  \left.\eu^{-s\vert\la\vert^2/2}\eu^{\la\cod+\la\st\co}\right\vert_{\la=0},\nonumber
\eea
for $n,m=0,1,2,\dots$. 
Beyond the \emph{value} $\Ord_s\co^n(\cod)^m$, our point of interest is the
\emph{scheme} (\ref{mix}) of ordering:
\beq\label{Ord_prod}
\Ord_s\co^n(\cod)^m=\sum_{P=1}^{(n+m)!}w_P [\co^n(\cod)^m]_P,
\eeq
where $w_P\geq0$ is the weight of the $P$th permutation, the weights are normalized;
$[\co^n(\cod)^m]_P$ stands for the $P$th permutation of the factors. 

We shall consider $\Ord_s\co^n(\cod)^m$ for $n=m=1$ first. If we wish to
bring it into the form on the r.h.s. of (\ref{Ord_prod}), we get the following
unique result:
\beq
\Ord_s\co\cod=\frac{1+s}{2}\cod\co+\frac{1-s}{2}\co\cod.
\eeq
As expected, $s\in[-1,1]$ interpolates between normal and anti-normal orderings. 
Next, let us take $n=2,m=1$. It is easy to derive, e.g.,  the following two forms (out of
an infinite family):
\beq
\Ord_s\co^2\cod=\cod\co^2+(1-s)\co=\co^2\cod-(1+s)\co.
\eeq
Using the identity $[\co,\cod]=1$, we could bring these two to two different forms, resp.,  
both of them conform with the r.h.s. of  (\ref{Ord_prod}). But the issue grows  unmanageable for 
higher powers $n,m$ unless we find a systematic way. Let us use 
our concept of orderings and the GTW (Sec. \ref{GWT}).

Inserting (\ref{s_T_equiv}) into (\ref{ord_prod_derivs}),  we obtain the general
expression
\bea\label{s-prod}
&&\Ord_s\co^n(\cod)^m=\left.\Tim
\left(\int_0^1\!\!\!\!\co_\tau d\tau\right)^n
\left(\int_0^1\!\!\!\!\cod_\tau d\chi_\tau\right)^m\right\vert_{\co_\tau=\co}
\!\!\!\!\!\!\!\!\!\!
\\
&=&\!\!\!\int_0^1\!\!\!\!\!d{\tau_1}\!\!\dots\!\!\int_0^1\!\!\!\!\!d{\tau_n}
\!\!\int_0^1\!\!\!\!\! d\chi_{\si_1}\!\!\dots\!\!\int_0^1\!\!\!\!\! d\chi_{\si_m}
\left.
\!\Tim\co_{\tau_1}\!\dots\!\co_{\tau_n}\cod_{\si_1}\!\dots\!\cod_{\si_m}
\right\vert_{\co_\tau=\co},\nonumber
\eea
which is already conform with the structure on the r.h.s. of (\ref{Ord_prod}).
Here we work out the special case $n=2,m=1$  only:
\bea
\Ord_s\co^2\cod&=&\int_0^1\!\!\!\!\!d\chi_\tau
\!\!\int_0^1\!\!\!\!\! d\si_1\int_0^1\!\!\!\!\! d\si_2
\left.
\!\Tim\cod_\tau\co_{\si_1}\co_{\si_2}
\right\vert_{\co_\tau=\co}\nonumber\\
&=&w_1\cod\co\co + w_2\co\cod\co + w_3\co\co\cod,
\eea
with
\bea\label{w_123}
w_1&=&\int_0^1\!\!\!\!\!d\chi_\tau
\!\!\int_0^1\!\!\!\!\! d\si_1\int_0^1\!\!\!\!\! d\si_2
\theta(\tau-\si_1)\theta(\tau-\si_2)\nonumber\\
&=&\int_0^1\!\!\!\!\! d\chi_\tau\tau^2,\nonumber\\
w_2&=&2\int_0^1\!\!\!\!\!d\chi_\tau
\!\!\int_0^1\!\!\!\!\! d\si_1\int_0^1\!\!\!\!\! d\si_2
\theta(\si_1-\tau)\theta(\tau-\si_2)\nonumber\\
&=&2\int_0^1\!\!\!\!\!d\tau\!\!\int_0^\tau\!\!\!\!\! d\chi_\tau\tau,\nonumber\\
w_3&=&\int_0^1\!\!\!\!\!d\chi_\tau
\!\!\int_0^1\!\!\!\!\! d\si_1\int_0^1\!\!\!\!\! d\si_2
\theta(\si_1-\tau)\theta(\si_2-\tau)\nonumber\\
&=&2\int_0^1\!\!\!\!\!d\tau\!\!\int_0^\tau\!\!\!\!\! d\tau \chi_\tau.
\eea
The constraint $w_1+w_2+w_3=1$ is satisfied by construction.
The three integrals (\ref{w_123}) depend on the choice of $\chi_\tau$. 
If we take a simplest function $\chi_\tau=\tau^\kappa$ with 
$\kappa=\frac{1+s}{1-s}\geq1$, covering the values $s\in[0,1)$, we get
\beq
\Ord_s\cod\co^2=\frac{(1-s)\cod\co\co+(1-s^2)\co\cod\co+(1+s)^2\co\co\cod}{3+s}.
\eeq
Obviously, other functions $\chi_\tau$ would have resulted in 
different weights $w_1,w_2,w_3$, i.e., in different schemes (\ref{Ord_prod}) of
the same non-monomial s-ordering.

\section{Concluding remarks}
In 1950, Wick's theorem between normal and chronological orderings 
of relativistic quantum fields was invented for and became the robust 
tool of S-matrix theory of interactions. The present
work started from the recognition that Wick's theorem, at least
the bosonic one, is the specific case of a simple relationship
between operator orderings in general. That is 
our general Wick theorem. Wick's original contraction is
replaced by the general contraction, a form which is fairly 
straightforward and transparent.
A little more involved, yet plausible, is the proposed definition of operator
orderings. It covers all common orderings, typical in quantum optics and 
quantum field theory, and opens a wide perspective toward new ones. 
The general Wick theorem clarifies the universal structure of orderings.
Contrary to naive expectations, the general Wick theorem
seems to hold for monomial orderings (permutations) only. Fortunately,
we could show that s-orderings of Cahill and Glauber (including the 
Weyl--Wigner ordering) are reductions of redundant monomial path orderings,
hence the general Wick theorem extends for them. 

Simple form of the general Wick theorem facilitates analytic calculations. 
We showed plain
explanatory applications in forced quantum mechanical motion,
a short derivation of the normal-ordered squeezing operator
and a less trivial application to elucidate s-orderings of Cahill and Glauber. 
The power of general orderings and their Wick theorem would get 
confirmed further by their future non-trivial utilizations.

It should be admitted on one hand that our proof of the general Wick theorem 
is tentative. On the other hand, more important is the non-triviality of this proof 
which tells us about the theorem's depths unless 
a simpler proof strategy pops up in the future. 

The author thanks the Hungarian Scientific Research Fund under Grant No. 124351, 
and the EU COST Action CA15220 for support. He is also grateful for the anonymous
adjudicator of JPA for important criticisms on the previous version of this work.


\vfill

\begin{thebibliography}{99}
\bibitem{Wey27} H. Weyl, Z. Phys. {\bf 46}, 1 (1927)
\bibitem{Wig32} E.P. Wigner, Phys. Rev. {\bf 40}, 749 (1932)
\bibitem{HouKin49} A. Hourlet and A. Kind, Helv. Phys. Acta {\bf 22}, 319 (1949)
\bibitem{Dys49} F.J. Dyson, Phys. Rev. {\bf 75}, 486 (1949)
\bibitem{Wic50} G.C. Wick, Phys. Rev. {\bf 80}, 268 (1950)
\bibitem{CahGla69} K.E. Cahill and R.J. Glauber, Phys. Rev. {\bf 177}, 1857 (1969)
\bibitem{Lee95} H.W. Lee, Phys. Rep.  {\bf 259} 147 (1995)
\bibitem{Choetal85} K. Chou, Zh. Su, B. Hao, and L. Yu, Phys. Rep. {\bf 118} 1 (1985)
\bibitem{BCH} J. Campbell, Proc. Lond. Math. Soc. {\bf 28}, 381 (1897);
                             H. Baker, Proc. Lond. Math. Soc. {\bf 34}, 347 (1902);
                             F. Hausdorff, Ber. Verh. Saechs. Akad. Wiss. Leipzig {\bf 58}, 19 (1906)
\bibitem{Fanetal} Fan Hong-Yi Fan  and Ruan Tu-Nan, Commun. Theor. Phys. (Beiging) {\bf 2}, 1289 (1983);
                                   Fan Hong-Yi, H.R. Zaidi and J.R. Klauder, Phys. Rev. {\bf D35}, 1831 (1987);
                                   A. W\"unsche, J. Optics B: Quantum Semiclass. Opt. {\bf 1}, R11 (1999).   
\end{thebibliography}
\end{document}